# МЕТОДЫ РЕШЕНИЯ КРАЕВЫХ ЗАДАЧ В ОБЛАСТЯХ С НЕСВЯЗНОЙ ГРАНИЦЕЙ

Трайтак С.Д.

*Институт прикладной механики РАН, г. Москва, Россия*

## РЕЗЮМЕ

На примере внешней задачи Дирихле для уравнения Лапласа в $\mathbf{R}^3$ проведен сравнительный анализ существующих аналитических методов решения краевых задач в средах, содержащих $N$ непересекающихся включений. Исследована область применимости метода отражений и метода изображений. Доказана теорема сложения для неприводимых декартовых тензоров с помощью которой решение данной краевой задачи сведено к решению бесконечной системы линейных алгебраических уравнений. Показано, что к той же системе приводит и известный метод индуцированных сил.

## 1. ВВЕДЕНИЕ

Краевые задачи для уравнений с частными производными в областях с несвязной границей играют важную роль в теории многих механических и физико-химических процессов, происходящих в гетерогенных средах. Среди них особую теоретическую и практическую значимость имеют краевые задачи теории переноса в средах с неподвижными включениями[1] [1-6]. В разные годы был предложен целый ряд как численных, так и аналитических методов решения этих краевых задач. Однако вплоть до настоящего времени все еще не определена область применимости данных методов и не выявлена их взаимосвязь.

Краткий обзор существующих численных методов решения задачи Дирихле в областях с несвязной границей для уравнения Пуассона дан в работе [7], кроме того, эффективные алгоритмы решения данной задачи, основанные на методе случайных блужданий по границе, подробно изложены в работах [8,9].

Настоящая статья посвящена строгому рассмотрению известных аналитических методов решения краевых задач в областях с несвязной границей с целью определения области применимости данных методов, а также установления их взаимной связи. Для определенности и простоты изложения здесь мы будем рассматривать лишь решения краевых задач для уравнения Лапласа.

К сожалению, в литературе (особенно посвященной приложениям) существует некоторое расхождение в определениях, поэтому вначале мы дадим некоторые определения, полезные для дальнейшего. Прежде всего, отметим, что для наших целей нет необходимости различать линейно связные и связные множества, поэтому под связными будем понимать линейно связные множества. Чтобы подчеркнуть особенности понятия многосвязности в трехмерном случае по аналогии с двумерным случаем иногда заменяют замкнутые кривые замкнутыми поверхностями и дают следующее определение [10].

---

[1] Отметим, что имея ввиду многочисленные приложения в теории теплопроводности, диффузии и гидродинамического взаимодействия, включения мы будем также называть частицами.

**Определение 1.** *Трехмерная область* $\Omega$ *называется пространственно односвязной, если любая замкнутая поверхность* $S \subset\subset \Omega$ *может быть непрерывно деформирована в точку* $P \in \Omega$, *в противном случае область называется пространственно многосвязной.*

Данное определение нам представляется не вполне удачным, так как в нем не делается разницы между топологически различными трехмерными областями, например, областями содержащими вырезы в виде сфер и торов. Поэтому мы дадим следующее

**Определение 2.** *Трехмерная область* $\Omega^* \subset \mathbf{R}^3$ *называется областью с несвязной границей, состоящей из* $N+1$ *компонент связности, если ее дополнение до* $\mathbf{R}^3$ *состоит из* $N+1$ *непересекающихся односвязных континуумов* $\overline{\Omega}_0$, $\overline{\Omega}_1$, $\overline{\Omega}_2$, ..., $\overline{\Omega}_N$, *причем* $\overline{\Omega}_0$ *содержит бесконечно удаленную точку* $\{\infty\}$.

В данной работе, для краткости, несвязную границу, состоящую из $N$ компонент связности будем называть просто $N$-связной границей. Отметим, что любой из континуумов $\overline{\Omega}_i$ ($i=\overline{1,N}$) может состоять только из одной точки. Важно подчеркнуть, что само множество $\Omega^*$ является не только связным, но и односвязным множеством. Определение 2 обобщает известное определение, данное в двумерном случае Векуа [11]. В частности по Векуа, при $N=0$, область $\Omega^*$ является односвязной с односвязной границей, а если $N=1$, то область $\Omega^*$ будет односвязной с двусвязной границей.

Однако в приложениях области с $N=1$ обычно считаются областями с односвязной границей. Последнее можно привести в соответствие с Определением 2, если добавить бесконечно удаленную точку $\{\infty\}$ к $\mathbf{R}^3$ и топологизировать пространство $\mathbf{R}^3 \cup \{\infty\}$ следующим образом. Любое множество $\Omega \subset \mathbf{R}^3 \cup \{\infty\}$ называется открытым, если оно является открытым подмножеством $\mathbf{R}^3$ или, если $\Omega = \left(\mathbf{R}^3 \setminus K\right) \cup \{\infty\}$, где $K \subset \mathbf{R}^3$ - некоторый компакт. Ясно, что стереографическая проекция осуществляет гомеоморфное отображение $\mathbf{R}^3 \cup \{\infty\}$ на единичную сферу в $\mathbf{R}^4$. Таким образом, мы получаем компактное топологическое пространство, которое называют одноточечной компактификацией $\mathbf{R}^3$ [12]. Учитывая данное обстоятельство в дальнейшем мы будем называть область $\Omega^*$ - односвязной областью с $N$-связной границей.

Рассмотрим область $\Omega^* = \mathbf{R}^3 \setminus \overline{\Omega}$, где $\overline{\Omega} = \Omega \cup \partial\Omega$ и $\Omega = \bigcup_{i=1}^{N} \Omega_i$. Мы полагаем, что $\overline{\Omega}_i$ - не пересекающиеся континуумы, тогда очевидно, что вся граница $\partial\Omega = \bigcup_{i=1}^{N} \partial\Omega_i$ является несвязным множеством, состоящим из $N$ связных компонент.

Краевые задачи для $N$-связной области $\Omega$ распадаются на соответствующие задачи в каждой связной компоненте $\Omega_i$ и поэтому не представляют особого интереса [13]. В связи с этим в дальнейшем мы рассматриваем лишь краевые задачи поставленные на $\Omega^*$.

Адекватно описать краевые задачи на множестве $\Omega^*$ можно лишь с помощью понятия дифференцируемого многообразия. Удобнее всего структуру гладкого ориентируемого многообразия на $\Omega^*$ можно ввести, рассматривая ориентируемый атлас $\left(\Omega^*, \varphi_i\right)$ ($i=\overline{1,N}$), где отображения $\varphi_i \in C^\infty\left(\Omega^*\right)$. Мы также положим, что $\partial\Omega \in C^2$, хотя большинство результатов, рассмотренных в статье, верны и для более общего случая поверхностей Ляпунова $C^{1,\lambda}$.

## 2. СВЕДЕНИЕ К ИНТЕГРАЛЬНЫМ УРАВНЕНИЯМ

В качестве простого, но важного для приложений примера будем рассматривать случай трехмерной внешней задачи Дирихле для уравнения Лапласа

$$\nabla^2 u = 0 \text{ в } \Omega^* \subset \mathbf{R}^3, \tag{1}$$

$$u(P)\big|_{\partial\Omega_i} = f_i(P_i) \tag{2}$$

с условием регулярности на бесконечности

$$u(P) = O\big(1/r(P_i, P)\big) \text{ при } r(P_i, P) \to \infty, \tag{3}$$

где $i=\overline{1,N}$ и $f_i(P_i) \in C(\partial\Omega_i)$; $r(P_i, P)$ - евклидово расстояние между текущей точкой $P \in \Omega^*$ и $P_i \in \partial\Omega_i$. Таким образом, мы имеем дело с внешней задачей Дирихле (1)-(3) в неограниченной односвязной области $\Omega^*$ с $N$-связной границей. Основная трудность при решении поставленной задачи состоит в невозможности выбора глобальных координат для всех связных компонент многообразия $\Omega^*$ если $N > 2$.

При решении аналогичной задачи о дифракции на двух рассеивающих телах Шварцшильд [14] предложил метод, основанный на функции Грина. Предполагая известной функцию Грина внешней задачи для каждой компоненты границы, он свел поставленную краевую задачу к системе из двух интегральных уравнений Фредгольма II рода. Шварцшильдом была доказана сходимость решения этой системы методом итераций. Отметим, что для случая, когда область $\Omega^*$ представляет собой внешность шаров, метод Шварцшильда можно свести к бесконечной системе линейных алгебраических уравнений совпадающей с системой получающейся с помощью прямого метода переразложения (см. ниже пункт 6). Аналогичную краевую задачу в теории упругости с помощью метода интегральных уравнений изучал также Михлин и Соболев [15].

Хорошо известно, что задача Дирихле (1)-(3) не имеет более одного решения [16]. Задачи Дирихле и Неймана для трехмерных многосвязных областей были впервые исследованы Гюнтером в монографии [13], где он свел их к решению интегрального уравнения Фредгольма II рода на границе области. Эти результаты в последующем были обобщены на случай смешанных краевых задач Дирихле-Неймана [17]. Однако, полученные интегральные уравнения, в отличие от случая области с односвязной границей, не являются однозначно разрешимыми и, следовательно, не могут быть решены прямым численным обращением соответствующих интегральных операторов [18].

В этой связи Крутицкий предложил разрешимое интегральное уравнение [18], причем им была рассмотрена смешанная краевая задача Дирихле-Неймана как в ограниченных так и в неограниченных трехмерных областях с многосвязной границей. Для простоты здесь мы представим лишь результаты для задачи Дирихле.

Рассмотрим односвязную область Дирихле $\Omega^*$ ограниченную простыми гладкими непересекающимися поверхностями $\partial\Omega_i$ ($i=\overline{1,N}$) такими, что $\partial\Omega\in C^{2,\lambda}$, $\lambda\in(0,1]$. Учитывая свойства потенциала двойного слоя ($\propto 1/r^2$ при $r\to\infty$), ищем решение поставленной краевой задачи в виде потенциалов двойного и простого слоя

$$f(P)\in C^{1,\lambda}(\partial\Omega),$$

(4)

$$u(P)=\frac{1}{4\pi}\int\limits_{\partial\Omega} dS_{P'}\mu(P')\frac{\partial}{\partial\mathbf{n}}\left[\frac{1}{r(P',P)}\right]+\frac{1}{4\pi}\sum_{j=1}^{N}\int\limits_{\partial\Omega_j} dS_{P'}\mu(P')\frac{1}{r(P_j,P)},$$

(5)

где $\mathbf{n}$ - внутренняя к $\Omega^*$ нормаль в точке $P'\in\partial\Omega$; $P_j\in\Omega_j$; ($j=\overline{1,N}$) произвольные фиксированные точки. Очевидно, что функция $u(P)$ является гармонической и удовлетворяет условию регулярности на бесконечности. Подстановка представления (5) в граничное условие (2) дает требуемое интегральное уравнение

$$\frac{1}{2}\mu(P)+\frac{1}{4\pi}\int\limits_{\partial\Omega} dS_{P'}\mu(P')\frac{\partial}{\partial\mathbf{n}}\left[\frac{1}{r(P',P)}\right]+$$

$$+\frac{1}{4\pi}\sum_{j=1}^{N}\int\limits_{\partial\Omega_j} dS_{P'}\mu(P')\frac{1}{r(P_j,P)}=f(P),\ P\in\partial\Omega.$$

(6)

Отметим, что второй интеграл здесь несобственный. В работе [18] была доказана разрешимость этого уравнения, и, таким образом, (6) может быть решено с помощью известных численных методов решения уравнений Фредгольма.

## 3. МЕТОД ОТРАЖЕНИЙ

Так называемый метод выметания Пуанкаре-Перрона применим при весьма общих предположениях о границе области и краевых условиях на ее границе [20], однако его трудно реализовать при решении прикладных задач. В связи с этим для получения приближенного аналитического решения внешней задачи Дирихле (1)-(3) в области с $N$-связной границей $\Omega^*\subset\mathbf{R}^3$ применяют идейно близкий к нему метод отражений. Данный метод очень часто используется в различных приложениях (см., например, работы [21-23], [19,24,25]). В классической книге Хаппеля и Бреннера [23] на стр.272 по поводу метода отражений отмечается: "Этот метод используется и в данной главе, хотя необходимо отметить, что до сих пор нет строгого доказательства сходимости итерационного процесса к искомому решению. Поэтому в настоящее время приходится довольствоваться ограниченными эмпирическими свидетельствами в пользу метода. Так, метод дает согласие с точным результатом Стимсона и Джефри для осесимметричной задачи о двух сферах [26]; в некоторых других случаях имеется согласие с существующими экспериментальными данными." В этой связи важно отметить, что до сих пор вопрос о сходимости метода отражений в $\mathbf{R}^3$ и его связи с другими подходами во многом остается открытым, поэтому здесь мы остановимся на нем подробно.

Идея метода отражений[2] состоит в следующем: используя известные решения в областях с односвязной границей $D_i \subset \mathbf{R}^3 \setminus \bar{\Omega}_i$ ($i=\overline{1,N}$) строится итерационная процедура для определения решения в области со сложной $N$-связной границей $\Omega^* = \bigcap_{i=1}^{N} D_i$ .

Построение приближений с помощью метода отражений хорошо известно (см., например, [20]) и приводит к $m$-му приближению вида

$$v_0(P) \equiv 0,$$

$$v_m(P) = v_{m-1}(P) + \sum_{k=1}^{N} u_k^{(m)}(P), \ (m \ge 1)$$

(7)

или

$$v_m(P) = \sum_{l=0}^{m} \sum_{k(\ne i)=1}^{N} u_k^{(l)}(P).$$

(8)

По построению эта функция является гармонической на $\Omega^*$, причем на каждой компоненте границы $\partial\Omega_i$ для функций $u_i^{(m)}(P)$ мы имеем равенство

$$u_i^{(m)}\Big|_{\partial\Omega_i} = f_i - v_{m-1}\big|_{\partial\Omega_i}, \ \ i=\overline{1,N}, \ m=\overline{0,\infty}.$$

(9)

Таким образом, ограничение функции $u_i^{(m)}(P)$ на $\partial\Omega_i$ является отклонением в граничном условии (2). Ясно, что если

$$\left\| u_i^{(m)}\Big|_{\partial\Omega_i} \right\|_\infty \to 0,$$

тогда решение задачи (1)-(3) дается формулой

$$u(P) = \lim_{m\to\infty} v_m(P) = \sum_{l=0}^{\infty}\sum_{i=1}^{N} u_i^{(l)}(P).$$

(10)

Для последовательности приближений $\{v_m(P)\}$ (дающейся формулой (8)) имеет место

**Утверждение 1.** Если $f_i > 0$, то функции $v_{2p}(P)$ ($v_{2p+1}(P)$), где $p=\overline{0,\infty}$ являются верхней (нижней) оценками точного решения $u(P)$, т.е. везде на $\bar{\Omega}^*$ справедливы неравенства

$$v_{2p}(P) > u(P),$$

(11)

$$v_{2p+1}(P) < u(P).$$

(12)

---

[2] В математических работах посвященных решению двумерных задач этот метод называют методом Шварца [19,26].

**Доказательство.** Легко видеть, что на $\bar{\Omega}^*$ верно неравенство

$$u_k^{(0)}(P) > 0$$

и, следовательно, в соответствии с методом отражений, имеют место следующие неравенства:

$$v_{2p}\Big|_{\partial\Omega_i} = f_i + \sum_{k(\neq i)=1}^{N} u_k^{(2p)}\Bigg|_{\partial\Omega_i} > f_i, \; i = \overline{1,N},$$

(13)

$$v_{2p+1}\Big|_{\partial\Omega_i} = f_i + \sum_{k(\neq i)=1}^{N} u_k^{(2p+1)}\Big|_{\partial\Omega_i} < f_i, \; i = \overline{1,N}.$$

(14)

Отсюда с помощью слабого принципа максимума для гармонических функций [15] получаем желаемые оценки (11) и (12).

Важно отметить, что доказанное следствие позволяет находить соответствующие двусторонние оценки в теории диффузионно-контролируемых процессов, которые ранее получали с помощью вариационных методов (см. работы [28,29] и ссылки в них).

Для случая когда $\partial\Omega \in C^{1,\lambda}$ Голузин доказал сходимость метода отражений для случая двух включений произвольной формы [30].

Здесь мы дадим более конструктивное доказательство теоремы о сходимости метода отражений, ограничив себя важным для приложений частным случаем, когда области $\Omega_i$ представляют собой открытые шары радиуса $R_i$ центры которых находятся в точках $P_i^{(0)}$. Это позволит нам оценить область применимости метода отражений в явном виде для общего случая $N > 2$.

Докажем вначале полезную оценку

**Лемма 1.** *Пусть гармоническая функция* $u : D \to \mathbf{C}$ *(где* $D \subset \mathbf{R}^3 \setminus \bar{\Omega}$ *и* $\Omega \subset \mathbf{R}^3$ *- ограниченная область) удовлетворяет условию* $|u(P')| \le C$ *для* $P' \in \partial\Omega$. *Рассмотрим гармоническую функцию* $v : D \to \mathbf{R}$, *такую, что* $v(P') = C$ *для* $P' \in \partial\Omega$. *Тогда имеет место оценка* $|u(P)| \le v(P)$ *для всех* $P \in D$.

**Доказательство.** Представляя $u(P)$ с помощью функции Грина $G(P,P')$ задачи Дирихле в области $D$ [16] имеем

$$|u(P)| = \frac{1}{4\pi}\left|\int_{\partial\Omega} u(P')\frac{\partial}{\partial\mathbf{n}}G(P,P')dS\right| \le \frac{1}{4\pi}\int_{\partial\Omega} |u(P')|\frac{\partial}{\partial\mathbf{n}}G(P,P')dS \le$$

$$\le \frac{1}{4\pi}\int_{\partial\Omega} C\frac{\partial}{\partial\mathbf{n}}G(P,P')dS = v(P).$$

Здесь мы учли, что $\dfrac{\partial}{\partial\mathbf{n}}G(P,P') > 0$ на $D$ [16].

Используя тот факт, что $f_i(P_i) \in C(\partial\Omega_i)$ имеем оценку

$$\left|u_k^{(0)}(P_k)\right| \le a_k = \max_{\partial\Omega_k}\left|f_k(P)\right| \text{ на } \partial\Omega_k, \; k = \overline{1,N}.$$

(15)

Вводя вспомогательные гармонические функции $v_k^{(0)}(P) = a_k R_k / r_k$, где $r_k = r\left(P, P_k^{(0)}\right)$ с помощью леммы 1 получаем

$$\left|u_k^{(0)}(P)\right| \le a_k R_k / r_k \text{ на } D_k,\ k = \overline{1,N}.$$

Очевидно, что

$$\max_{\partial\Omega_k} \frac{R_k}{r_k} = \frac{R_k}{L_{ik} - R_i} = b_{ki},\ k(\ne i) = \overline{1,N},$$

где $L_{ik}$ - расстояние между центрами $i$-го и $k$-го шаров, причем в случае непересекающихся шаров ясно, что всегда $b_{ki} < 1$ (см. рис.1).

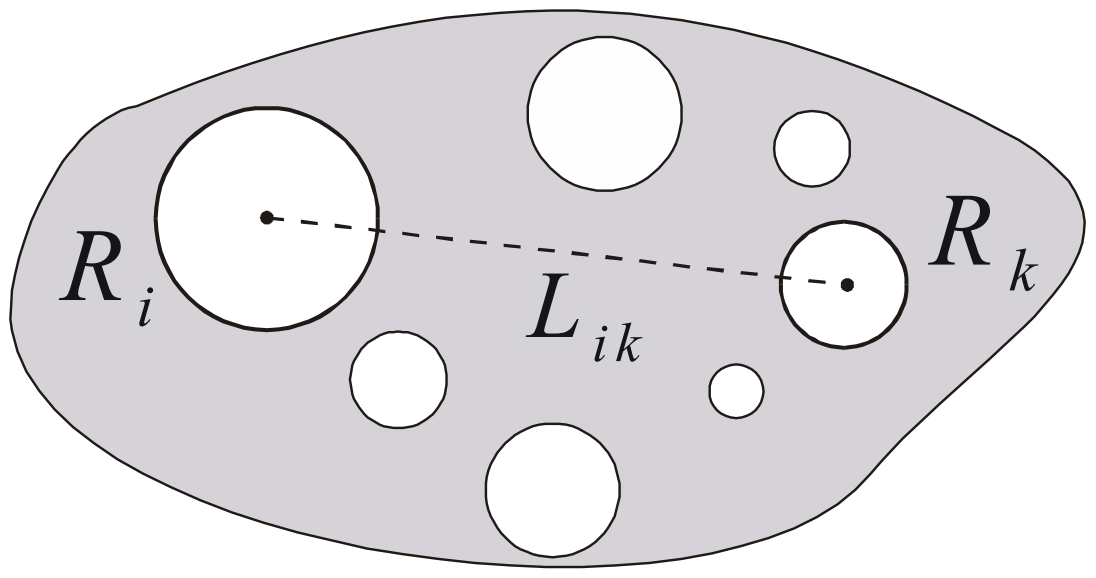

Рис.1. Случай неограниченной области $\Omega^*$ с $N$-связной границей, которую составляют непересекающиеся сферы.

Воспользовавшись неравенством (15), отсюда получаем

$$\left|u_k^{(0)}(P_i)\right| \le a_k b_{ki} \text{ на } \partial\Omega_i,\ k(\ne i) = \overline{1,N}.$$

Более того, очевидно, что

$$\left|u_k^{(0)}(P_i)\right| \le a^{(0)} b_{ki} \text{ на } \partial\Omega_i,\ k(\ne i) = \overline{1,N},$$

(16)

где $a^{(0)} = \max\{a_1, a_2, ..., a_N\}$. При помощи условия (9) и неравенства (16), имеем оценку

$$\left|u_i^{(1)}(P_i)\right| \le \sum_{k(\ne i)=1}^{N} \left|u_k^{(0)}(P_i)\right| \le a^{(0)} b^{(0)} \equiv a^{(1)}, \text{ на } \partial\Omega_i,\ i = \overline{1,N},$$

где $b^{(0)} = \max_{i=\overline{1,N}} \sum_{k(\ne i)=1}^{N} b_{ki}$. Аналогичные рассуждения приводят на сфере $\partial\Omega_i$ к неравенству

$$\left|u_k^{(1)}(P_i)\right| \le a^{(1)} b^{(0)},$$

а, проведя оценки вплоть до числа $m$, находим

$$\left|u_i^{(m)}(P_i)\right| \le a^{(m)} b^{(0)},$$

где $a^{(m)} = a^{(m-1)} b^{(0)}$.

Таким образом, окончательно мы получаем оценку

$$\left|u_i^{(m)}(P_i)\right| \le a^{(0)} \left[b^{(0)}\right]^m \text{ на } \partial\Omega_i,\ i = \overline{1,N}.$$

(17)

Отсюда следует, что при условии

$$b^{(0)} < 1$$

(18)

или, более подробно,

$$\max_{i=1,N} \sum_{k(\neq i)=1}^{N} \left( \frac{\varepsilon_{ki}}{1-\varepsilon_{ik}} \right) < 1,$$

(19)

где $\varepsilon_{ki} = R_k / L_{ik}$, последовательность $\left\{ u_i^{(m)}(P) \right\}$ равномерно сходится к нулю на границе $\partial\Omega^*$, а, вследствие известной теоремы Харнака [16], и везде в области $\overline{\Omega^*}$.

Совершенно аналогично можно показать, что необходимым условием сходимости метода отражений является условие

$$\min_{i=1,N} \sum_{k(\neq i)=1}^{N} \left( \frac{\varepsilon_{ki}}{1+\varepsilon_{ik}} \right) < 1.$$

(20)

Таким образом доказана

**Теорема 1.** *Для того чтобы последовательность* $\left\{ v_m(P) \right\}$ *равномерно сходилась на* $\overline{\Omega^*}$ *к решению задачи (1)-(3), должны выполняться достаточное (18) ((19)) и необходимое (20) условия.*

Пусть, например, $N = 2$, тогда условие сходимости (19) принимает вид

$$L_{12} > R_1 + R_2 .$$

Другими словами, если две частицы не касаются друг друга, то, например, стационарную диффузию вне этих частиц всегда можно изучать методом отражений.

В частном случае одинаковых шаров имеем следующие достаточное и необходимое условия сходимости метода отражений:

$$\max_{i=1,N} \sum_{k(\neq i)=1}^{N} \left( \frac{\varepsilon_{ki}}{1-\varepsilon_{ki}} \right) < 1, \quad \min_{i=1,N} \sum_{k(\neq i)=1}^{N} \left( \frac{\varepsilon_{ki}}{1+\varepsilon_{ki}} \right) < 1.$$

(21)

С помощью этих двух неравенств нетрудно построить предельные кривые (точнее говоря конечную последовательность точек, зависящую от числа $N$), ограничивающие области заведомой сходимости и расходимости метода отражений соответственно. Этот вывод является принципиально важным, так как до настоящего времени считалось, что задача Дирихле для уравнения Лапласа (Стокса) разрешима методом отражений для любого числа частиц и только вычислительные трудности не позволяют осуществить решение для большого числа частиц [23].

На рис.2 показаны предельные кривые сходимости метода отражений для случая, когда $\Omega^*$ представляет собой внешность одномерной цепочки одинаковых сферических включений радиусом $R$ расположеных с шагом $a$. Видно, что, например, для $N = 100$ метода отражений неприменим при $a/R < 4{,}9$, а при $a/R < 9{,}4$ его применимость требует дополнительного обоснования.

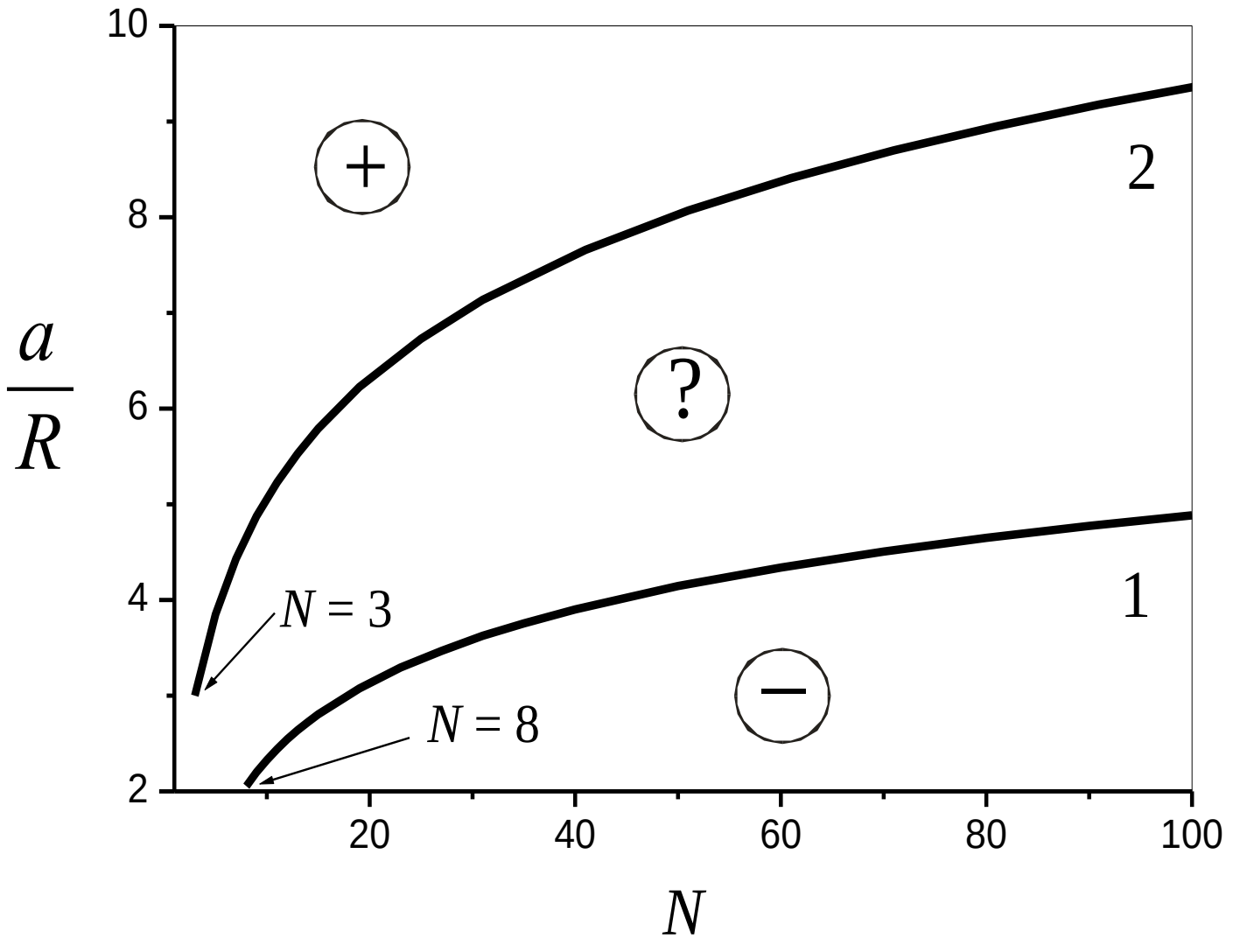

Рис.2. Предельные кривые сходимости метода отражений для одномерной цепочки одинаковых сферических включений. Кривая 1 - предельная кривая необходимого условия сходимости (20); . кривая 2 - предельная кривая достаточного условия сходимости (19); + - область заведомой сходимости; - - область заведомой расходимости; ? - область для которой необходимо дополнительное исследование сходимости.

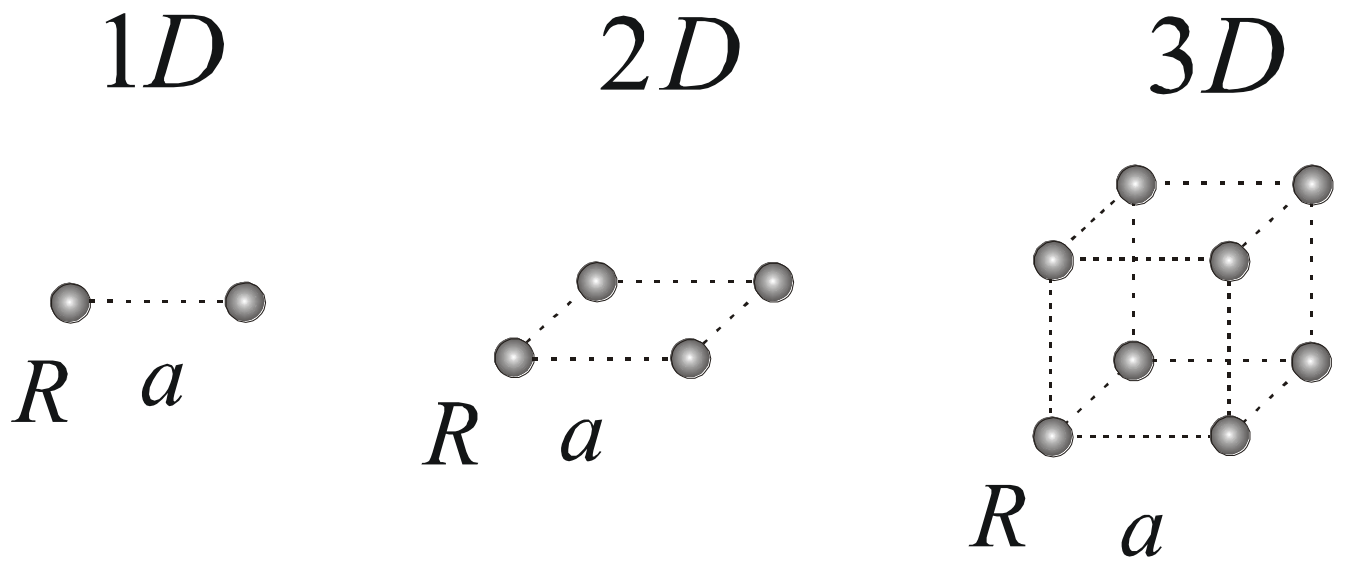

Рис.3. Кластеры одинаковых сферических включений, составляющих простые $D$ - мерные решетки.

Для внешности кластеров сферических частиц образующих двумерную и трехмерную решетку (см. рис.3) как видно из рис.4 область заведомой неприменимости метода отражений возрастает.

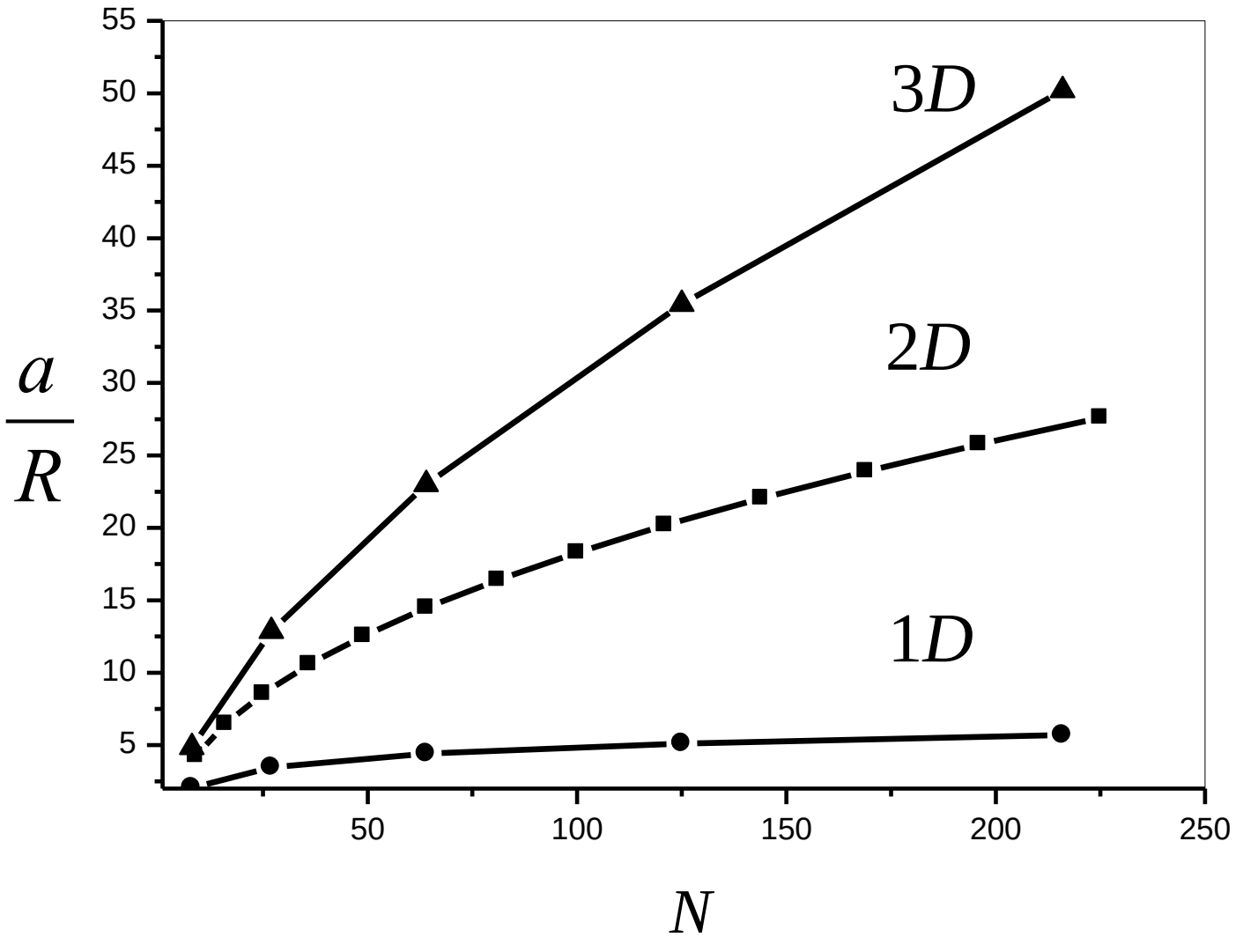

Рис.4. Предельные кривые расходимости метода отражений для случая простых $D$ - мерных решеток одинаковых сферических включений.

## 4. МЕТОД ИЗОБРАЖЕНИЙ

Метод изображений применим к решению задачи Дирихле в области $\Omega^*$ вне шаров $\overline{B}_i$. Обычно он формулируется для вспомогательных функций $u_i$ при следующих краевых условиях [31,32]

$$u_i\big|_{\partial B_i} = u_0^i = const,\ u_i\big|_{\partial B_j} = 0,\ j(\neq i) = \overline{1,N}\ .$$

(22)

Искомое решение соответствующей задачи Дирихле (1), (22), (3) дается суммой

$$u = \sum_{i=1}^{N} u_i\ .$$

(23)

Для простоты мы будем рассматривать задачу Дирихле лишь в области с двусвязной границей ( $N = 2$ ).

Без ограничения общности можно считать, что

$$u_i\big|_{\partial B_i} = 1,\ u_i\big|_{\partial B_j} = 0,\ j(\neq i) = 1,2\ .$$

(24)

В качестве нулевого приближения выберем гармоническую функцию, удовлетворяющую граничному условию на $\partial B_i$:

$$u_i^{(0)} = R_i / r_i\ ,$$

(25)

где $r_i$ - радиальная координата в локальной сферической системе координат связанной с $i$ -ой сферой (для $i = 1$ см. рис.5).

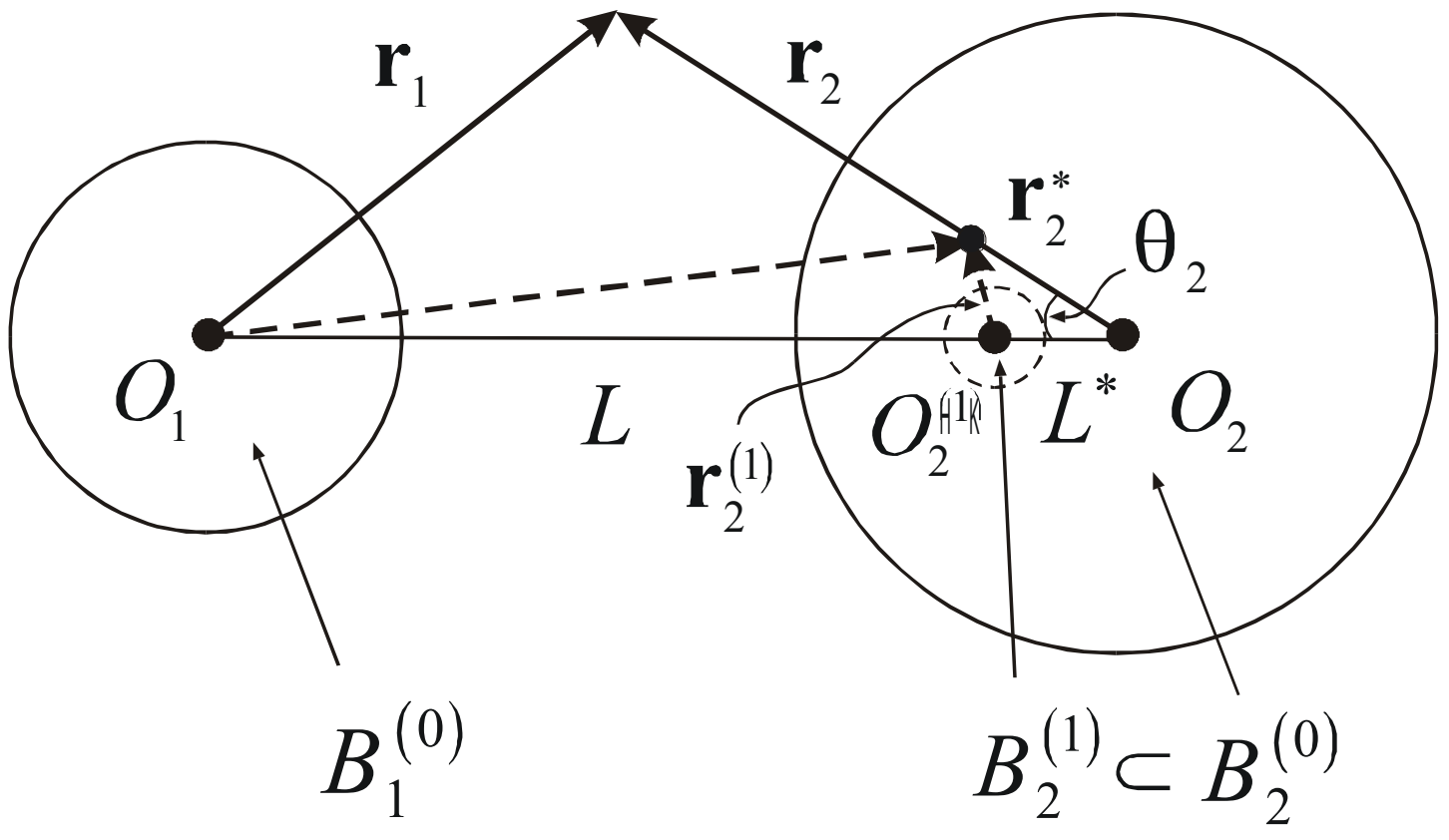

Рис.5. Схема метода изображений.

Функция (25) в локальной системе координат связанной с $j$-ой сферой запишется как

$$u_i^{(0)} = \frac{R_i}{\sqrt{r_j^2 + L^2 - 2r_j L \mu_j}},$$

(26)

где $\mu_i = \cos\theta_i$. Чтобы скомпенсировать влияние (26) на $j$-ой сфере, введем новую функцию $u_i^{(1)}$ так, что

$$\left(u_i^{(0)} + u_i^{(1)}\right)\Big|_{\partial B_j} = 0.$$

(27)

Согласно методу изображений $u_i^{(1)}$ определяется как преобразованная по Кельвину относительно $j$-ой сферы [16] функция взятая с противоположным знаком, т.е. [32]

$$u_i^{(1)} = -K_j \circ u_i^{(0)} = -\frac{R_i R_j}{L\sqrt{r_j^{*2} + L^{*2} - 2r_j^* L^* \mu_j}},$$

(28)

где $r_j^* r_j = R_j^2$, $L^* L = R_j^2$; $r_j^*$ и $L^*$ это образы точек $r_j$ и $L$ при преобразовании Кельвина. Отметим, что последовательность всех образов точки $r_i$ находится внутри рассматриваемых сфер. Легко заметить, что выражение (28) переписывается в виде

$$u_i^{(1)} = -\frac{R_i R_j}{L r_j^{(1)}}.$$

(29)

Таким образом, функция $u_i^{(1)}$ может быть представлена в форме потенциала точечного заряда: $-R_i R_j / L$ с полюсом в точке $O_j^{(1)}$ находящимся на расстоянии $L^*$ от $O_j$ (см. рис.5). Очевидно, что полюс $O_j^{(1)}$ является центром шара $\overline{B}_j^{(1)}$, получающегося из шара $\overline{B}_i$ преобразованием Кельвина относительно сферы $\partial B_i$, причем

$$\bar{B}_j^{(1)} = K_j \circ \bar{B}_i \subset \bar{B}_j .$$

Совершенно аналогично компенсируем изменение на $i$-ой сфере вызванное наличием введенного поля $u_i^{(1)}$ и т. д. Получающаяся при этом последовательность функций $\left\{u_i^{(m)}\right\}$ определяется рекуррентными формулами

$$u_i^{(2k+1)}\left(\mathbf{x}_j\right) = -K_j \circ u_i^{(2k)}\left(\mathbf{x}_j\right),\ k \geq 0;$$

$$u_i^{(2k)}\left(\mathbf{x}_i\right) = -K_i \circ u_i^{(2k-1)}\left(\mathbf{x}_i\right),\ k \geq 1.$$

(30)

**Теорема 2.** *Последовательность функций* $\left\{u_i^{(m)}\right\}$ *является сходящейся.*

Отметим, что обычно обоснование этого факта проводится непосредственно после вычисления явного вида членов последовательности $\left\{u_i^{(m)}\right\}$ и поэтому имеет довольно громоздкий вид [32]. Дадим здесь другое, более краткое доказательство данного утверждения.

**Доказательство.** Получающаяся последовательность компенсирующих функций $\left\{u_i^{(m)}\right\}$ представляет собой потенциалы точечных зарядов, полюса которых находятся в центрах шаров $\bar{B}_j^{(m)}$ ($i = 1, 2$) определяемых рекуррентной формулой

$$\bar{B}_i^{(0)} = \bar{B}_i ,$$

$$\bar{B}_j^{(m)} = K_i \circ \left(K_j \circ \bar{B}_i^{(m-1)}\right),\ m \geq 1,\ j(\neq i) = 1, 2 .$$

(31)

Легко видеть, что $\left\{\bar{B}_i^{(m)}\right\}$ это последовательность вложенных замкнутых шаров, т.е. $\bar{B}_i^{(m)} \subset \bar{B}_i^{(m-1)}$ ($m \geq 1$) радиусы которых стремятся к нулю. Из полноты метрического пространства ($\mathbf{R}^3, \rho$), где $\rho$ - евклидова метрика, следует существование единственной точки $\mathbf{x}_i^* \in \bigcap_{m=0}^{\infty} \bar{B}_i^{(m)}$ - полюса предельного потенциала точечного заряда. Таким образом, имеем

$$\lim_{k\to\infty} u_i^{(2k+1)} \propto \frac{1}{\left|\mathbf{r} - \mathbf{x}_i^*\right|}, \qquad \lim_{k\to\infty} u_i^{(2k)} \propto \frac{1}{\left|\mathbf{r} - \mathbf{x}_j^*\right|}.$$

Из единственности решения задачи Дирихле (1), (27), (3) следует, что последовательность $\{v_m\}$ $m$-частичных сумм вида

$$v_m = \sum_{k=0}^{m}\left[u_1^{(k)} + u_2^{(k)}\right]$$

(32)

совпадает с последовательностью $\{v_m\}$, получающейся в методе отражений. Отсюда вытекает, что метод изображений является частным случаем метода отражений, когда функции $f_i\left(P_i\right)$ в граничных условиях (2) тождественно равны некоторым постоянным.

Следовательно, решение вспомогательной задачи $u_i$ дается рядом

$$u_i = u_i^{(0)} - K_j \circ u_i^{(0)} + K_i \circ K_j \circ u_i^{(0)} - K_j \circ K_i \circ K_j \circ u_i^{(0)} + \ldots$$

(33)

или, учитывая равномерную сходимость (33), находим

$$u_i = \left(I - K_j\right) \circ \sum_{k=0}^{n} \left(K_i \circ K_j\right)^k \circ u_i^{(0)} = \left(I - K_j\right) \circ \left(I - K_i \circ K_j\right)^{-1} \circ u_i^{(0)},$$

(34)

где $I$ - тождественный оператор. Отметим, что равномерная сходимость как ряда (33) так и ряда $u_1 + u_2$ следует из теоремы 1. Этот факт может быть также доказан непосредственно с помощью свойств преобразования Кельвина.

Отличительной особенностью классического метода изображений является возможность обойтись без использования теорем сложения для гармонических функций. Соответствующие компенсирующие решения определяются лишь с помощью потенциалов точечных зарядов. Однако это упрощение приводит к следующим ограничениям метода изображений: а) метод применим лишь к краевым задачам для уравнения Лапласа, решения которого, как известно, инвариантны относительно преобразования Кельвина; б) область в которой решается задача Дирихле имеет вид только внешности непересекающихся сфер; в) решение возможно лишь в случае постоянных значений искомой функции на границе, либо условий, когда зависимость $f_i\left(P_i\right)$ индуцирована точечными зарядами, находящимися внутри сфер.

Важно отметить, что Голузин обобщил метод изображений на случай произвольных $f_i \in C\left(\partial\Omega_i\right)$, сведя задачу Дирихле к решению некоторой системы функциональных уравнений [30].

## 5. ОБОБЩЕННЫЕ ТЕОРЕМЫ СЛОЖЕНИЯ

В случае описания стационарной диффузии в гетерогенной среде локальная концентрация является функцией вида $u : \Omega^* \to \mathbf{R}_+$, а для того, чтобы вычислять ее значения и локальные потоки, необходимо ввести локальные координаты и рассмотреть функцию $u'\left(\mathbf{x}_i\right)$ связанную с каждой картой $\left\{\Omega^*, \varphi_i\right\}$:

$$u'\left(\mathbf{x}_i\right) = u\left(P\right) \text{ для } \mathbf{x}_i = \varphi_i\left(P\right) \text{ и любой точкой } P \in \Omega^*.$$

Для любой другой карты $\left\{\Omega^*, \varphi_j\right\}$ в одной и той же точке $P \in \Omega^*$ имеем

$$u'\left(\mathbf{x}_i\right) = u''\left(\mathbf{x}_j\right) \text{ для } \mathbf{x}_j = \varphi_j\left(P\right)$$

или

$$u''\left(\varphi_j\left(\varphi_i^{-1}\left(\mathbf{x}_i\right)\right)\right) = u'\left(\mathbf{x}_i\right),$$

(35)

т. е.функция $u''$ выражается через локальные координаты $\left\{\mathbf{x}_i\right\}$, причем функция $u'$ аналогично может быть выражена через локальные координаты $\left\{\mathbf{x}_j\right\}$.

Рассмотрим важный частный случай, когда $\overline{\Omega}_i = \overline{B}_{R_i}\left(\mathbf{x}_i^0\right)$ $(i = \overline{1, N})$ - замкнутые шары радиусами $R_i$, центры которых помещены в точки $\mathbf{x}_i^0$ некоторой трехмерной декартовой системы координат. В этом случае формулы координатных преобразований

$$\mathbf{x}_j = \varphi_j\left(\varphi_i^{-1}(\mathbf{x}_i)\right)$$

(36)

образуют группу движений в $\mathbf{R}^3$ и принимают простой вид

$$\mathbf{x}_j = \mathbf{O}_r \cdot \mathbf{x}_i + \mathbf{b}_{ij},$$

(37)

где $\mathbf{b}_{ij}$ - постоянные векторы трансляций, а $\mathbf{O}_r$ - матрица вращений. В частности для областей $D_i$ являющихся внешностями шаров $\bar{B}_{R_i}\left(\mathbf{x}_i^0\right)$, в качестве локальных удобно выбирать сферические координаты, связанные с $\bar{B}_{R_i}$: $\{\mathbf{x}_i\} = (r_i, \theta_i, \phi_i)$, где $\{r_i > 0,\, 0 < \theta_i \le \pi,\, 0 < \phi_i \le 2\pi\}$.

Важно отметить, что этот случай при $N = 2$ можно описать с помощью одной бисферической координатной системы $(\xi, \eta, \phi)$, т. е. атлас многообразия $\Omega^*$ состоит из одной карты $\left\{\Omega^*, \varphi\right\}$, где диффеоморфизм $\varphi^{-1} : \Pi \to \Omega^*$ определяется формулами

$$x = \rho\cos\phi, \quad y = \rho\sin\phi, \quad z = a\,\mathrm{sh}\xi/(\mathrm{ch}\xi - \cos\eta).$$

(38)

Здесь $\rho = a\sin\eta/(\mathrm{ch}\xi - \cos\eta)$, $a$ - параметр бисферических координат, а область определения $\Pi$ является открытым неограниченным параллелепипедом: $\left\{\xi \in (-\infty, \infty),\, \eta \in (0, \pi),\, \phi \in (0, 2\pi)\right\}$.

В работе [33] для внутренней задачи Дирихле в области между двумя неконцентрическими сферами была использована система сдвинутых внутренних координат, которая также позволяет описать обе связные компоненты границы с помощью одной координатной системы. Хорошо известно, что, используя преобразование Кельвина, это решение можно единственным образом отобразить в решение соответствующей внешней задачи Дирихле [16]. Отметим также, что как бисферические координаты, так и система сдвинутых внутренних координат [33] конформно эквивалентны сферической системе координат [34].

Предположим, что в локальных координатах $\{\mathbf{x}_i\}$ существуют решения уравнения (1) - $\left\{\psi_k^i(\mathbf{x}_i)\right\}\Big|_{k=0}^{\infty}$ (здесь и далее $k$ в общем случае является мультииндексом), образующие полную систему функций в гильбертовом пространстве $L_2(D_i)$ и поэтому

$$u'(\mathbf{x}_i) = \sum_{k=0}^{\infty} \alpha_k^i \psi_k^i(\mathbf{x}_i).$$

В других локальных координатах $\{\mathbf{x}_j\}$ мы, вообще говоря, имеем другую полную в $L_2(D_j)$ систему решений $\left\{\chi_k^j(\mathbf{x}_j)\right\}\Big|_{k=0}^{\infty}$ и, следовательно,

$$u''(\mathbf{x}_j) = \sum_{k=0}^{\infty} \alpha_k^j \chi_k^j(\mathbf{x}_j).$$

Таким образом, в пространстве $L_2(D_i \cap D_j)$ получаем связь

$$\chi_k^j\left(\mathbf{x}_j\right)=\sum_{p=0}^{\infty}U_{kp}^{ij}\psi_p^i\left(\mathbf{x}_i\right)$$

(39)

или, воспользовавшись соотношением (36), выражение (39) запишется в координатах $\left\{\mathbf{x}_i\right\}$ как

$$\chi_k^j\left(\varphi_j\left(\varphi_i^{-1}\left(\mathbf{x}_i\right)\right)\right)=\sum_{p=0}^{\infty}U_{kp}^{ij}\psi_p^i\left(\mathbf{x}_i\right).$$

(40)

Очевидно, что матрицы $\left(U_{kp}^{ij}\right)$ являются соответствующими коэффициентами Фурье функций $\chi_k^j$ по системе $\left\{\psi_k^i\left(\mathbf{x}_i\right)\right\}\Big|_{k=0}^{\infty}$, т.е.

$$U_{kp}^{ij}=\left\|\psi_p^i\left(\mathbf{x}_i\right)\right\|_{L_2\left(D_i\cap D_j\right)}^{-2}\int\chi_k^j\left(\varphi_j\left(\varphi_i^{-1}\left(\mathbf{x}_i\right)\right)\right)\psi_p^i\left(\mathbf{x}_i\right)d\mathbf{x}_i.$$

(41)

Как правило, при вычислении этих интегралов возникают трудности, поэтому случаи, когда матрица $\left(U_{kp}^{ij}\right)$ может быть найдена без этих вычислений, представляют большой практический интерес. Соотношение (39) с известной матрицей $\left(U_{kp}^{ij}\right)$ будем называть обобщенной теоремой сложения или формулой переразложения, а входящие в него матрицы $\left(U_{kp}^{ij}\right)$ - матрицами перехода или, для диффузионных задач, матрицами диффузионного взаимодействия.

Данное выше определение обобщает известный общий вид теорем сложения вытекающий из равенства, связывающего функции одинакового вида аргументы одной из которых преобразованы под действием трансляций [35]. Отметим также, что с другой стороны стандартная теорема сложения для сферических гармоник связана лишь с вращением [36], но не с трансляциями!

Ниже показано, что если известна соответствующая теорема сложения, то решение краевой задачи (1)-(3) может быть осуществлено с помощью решения соответствующей бесконечной системы линейных алгебраических уравнений (БСЛАУ). Этот метод решения задачи Дирихле назовем прямым методом переразложения.

## 6. СВЯЗЬ МЕТОДА ОТРАЖЕНИЙ С МЕТОДОМ ПЕРЕРАЗЛОЖЕНИЯ

Для простоты рассмотрим случай, когда области $\Omega_i$ однотипны, т. е. все $\Omega_i$ являются, например, шарами, сфероидами, эллипсоидами и другими областями границы которых образуют координатные поверхности систем в которых уравнение Лапласа (1) допускает $R$-разделение переменных [34]. Пусть $\left\{\psi_k\left(\mathbf{x}_i\right)\right\}\Big|_{k=0}^{\infty}\subset\ker\nabla^2$ - базис в гильбертовом пространстве $L_2\left(D_i\right)$ функций $\psi_k\left(\mathbf{x}_i\right)$ такой, что ограничение на $\partial\Omega_i$: $\left\{\psi_k\left(\mathbf{x}_i\right)\Big|_{\partial\Omega_i}\right\}\Big|_{k=0}^{\infty}$ в свою очередь является базисом в гильбертовом пространстве $L_2\left(\partial\Omega_i\right)$. Тогда элементы

последовательности $\left\{u_i^{(m)}\right\}$ возникающей при применении метода отражений в $D_i$ определяются как

$$u_i^{(m)} = \sum_{k=0}^{\infty} A_k^{i(m)} \psi_k(\mathbf{x}_i),$$

(42)

где $A_k^{i(m)}$ - соответствующие коэффициенты Фурье. Если воспользоваться этой общей формулой и теоремой сложения (39) в граничных условиях (2), то согласно методу отражений имеем

$$A_k^{i(0)} = \alpha_k^i,\ A_k^{i(m)} = -\sum_{j(\neq i)=1}^{N} \sum_{k=0}^{\infty} U_{kp}^{ij} A_p^{j(m-1)} \text{ для } m \geq 1,$$

(43)

где $\alpha_k^i = \left\| \psi_k \big|_{\partial\Omega_i} \right\|_{L_2(\partial\Omega_i)}^{-1} \int_{\partial\Omega_i} (f_i \psi_k)\big|_{\partial\Omega_i} dS_i$.

В соответствии с теоремой 1 решение задачи Дирихле (1)-(3) дается равномерно сходящимся рядом (10), который, с учетом (42) можно записать как

$$u(P) = \sum_{i=1}^{N} \sum_{k=0}^{\infty} \sum_{m=0}^{\infty} A_k^{i(m)} \psi_k(\mathbf{x}_i).$$

(44)

С другой стороны решение этой задачи можно искать в виде

$$u(P) = \sum_{i=1}^{N} \sum_{k=0}^{\infty} A_k^i \psi_k(\mathbf{x}_i).$$

(45)

Сравнение выражений (44) и (45) дает очевидную связь соответствующих коэффициентов

$$A_k^i = \sum_{m=0}^{\infty} A_k^{i(m)}.$$

Непосредственная подстановка разложения (45) в граничные условия (2) с использованием теоремы сложения (39) дает БСЛАУ для определения неизвестных коэффициентов $A_k^i$:

$$A_k^i = \alpha_k^i - \sum_{j(\neq i)=1}^{N} \sum_{p=0}^{\infty} U_{kp}^{ij} A_p^j.$$

(46)

Отсюда легко видеть, что формальное итерационное решение этой системы также дается формулами (43) и таким образом, нами доказано важное утверждение

**Теорема 3.** *Решение БСЛАУ (46) методом простой итерации эквивалентно определению коэффициентов $A_k^i$ с помощью метода отражений (43).*

В работе [37] на стр.72 метод отражений подвергается критике: "... этот метод представляется нам несколько громоздким, что в определенной степени объясняется непоследовательностью и даже искусственностью в математической постановке задач, которая осуществляется в данном подходе." Далее говорится о модифицированном методе отражений, который состоит в использовании общей теоремы сложения для регулярных шаровых функций, представленных в локальных сферических координатах. Однако в то время авторам была известна лишь осесимметричная теорема сложения и для решения задачи о гидродинамическом взаимодействии произвольно расположенных $N$ сфер в статье [37] была получена соответствующая теорема сложения в терминах декартовых неприводимых тензоров. С помощью этой теоремы сложения из граничных условий была найдена бесконечная система линейных алгебраических уравнений относительно неизвестных коэффициентов, что и составляет суть нового метода, предложенного в работе [37]. Ниже на стр.80 читаем: "Данная система уравнений, как нетрудно видеть, может быть решена методом последовательных приближений с точностью до любого порядка по параметру $h$." Здесь $h \equiv \varepsilon$. Однако, во-первых из доказанной выше теоремы следует, что решение методом последовательных приближений приводит к тому же самому результату, что и критикуемый авторами метод отражений. Во-вторых, из результатов пункта 3 следуют ограничения при применении метода итераций в зависимости от числа частиц и величины параметра $\varepsilon$. Следовательно, для того чтобы выйти за рамки метода отражений в общем случае необходимо решать БСЛАУ (46) методом, отличным от метода простых итераций.

Итерационное решение соответствующей БСЛАУ удобно реализовывать в матричной форме

$$(\mathbf{I}+\mathbf{U})\mathbf{A}=\mathbf{B}.$$

(47)

Здесь

$$\mathbf{I}=\begin{pmatrix}\mathbf{E} & & \mathbf{0}\\ & \ddots & \\ \mathbf{0} & & \mathbf{E}\end{pmatrix},\quad \mathbf{U}=\begin{pmatrix}\mathbf{0} & \mathbf{U}^{(1,2)} & \cdots & \mathbf{U}^{(1,N)}\\ \mathbf{U}^{(2,1)} & \mathbf{0} & & \mathbf{U}^{(2,N)}\\ \vdots & & \ddots & \vdots\\ \mathbf{U}^{(N,1)} & \mathbf{U}^{(N,2)} & \cdots & \mathbf{0}\end{pmatrix},$$

$$\mathbf{A}=\begin{pmatrix}\mathbf{A}^{(1)}\\ \vdots\\ \mathbf{A}^{(N)}\end{pmatrix},\quad \mathbf{B}=\begin{pmatrix}\mathbf{B}^{(1)}\\ \vdots\\ \mathbf{B}^{(N)}\end{pmatrix},$$

где $\mathbf{E}$ - бесконечная единичная матрица, $\mathbf{U}^{(i,j)}$ - соответствующие матрицы взаимодействия частиц $i$ и $j$, $\mathbf{A}^{(i)}$ и $\mathbf{B}^{(i)}$ - последовательности неизвестных коэффициентов и свободных членов, соответственно. При условии, что решение матричного уравнения (47) существует имеем

$$\mathbf{A}=(\mathbf{I}+\mathbf{U})^{-1}\mathbf{B}.$$

С другой стороны хорошо известно, что если выполняется условие

$$\|\mathbf{U}\|<1$$

(48)

(либо (19)), то решение может быть найдено итерациями, т.е.

$$\mathbf{A}=\left(\sum_{k=0}^{\infty}(-1)^k\mathbf{U}^k\right)\mathbf{B}=\sum_{k=0}^{\infty}\mathbf{A}_k .$$

(49)

Отсюда видно, что

$$\mathbf{A}_k=-\mathbf{U}\mathbf{A}_{k-1}.$$

Важно отметить, что для случая диффузии численные расчеты показали, что достаточное условие (19) является менее сильным, чем условие (48) и, более того,

БСЛАУ для конечного числа частиц различного радиуса является квазирегулярной (см. книгу [38]).

Для того чтобы выходить за пределы метода итераций, можно, в частности, осуществлять частичное суммирование в формальном итерационном ряде. Последнее удобно проводить с помощью следующего диаграммного представления:

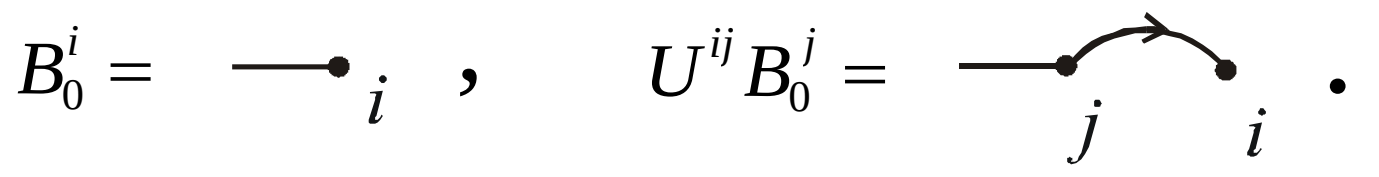

Тогда решение БСЛАУ (47) формально можно записать в виде ряда

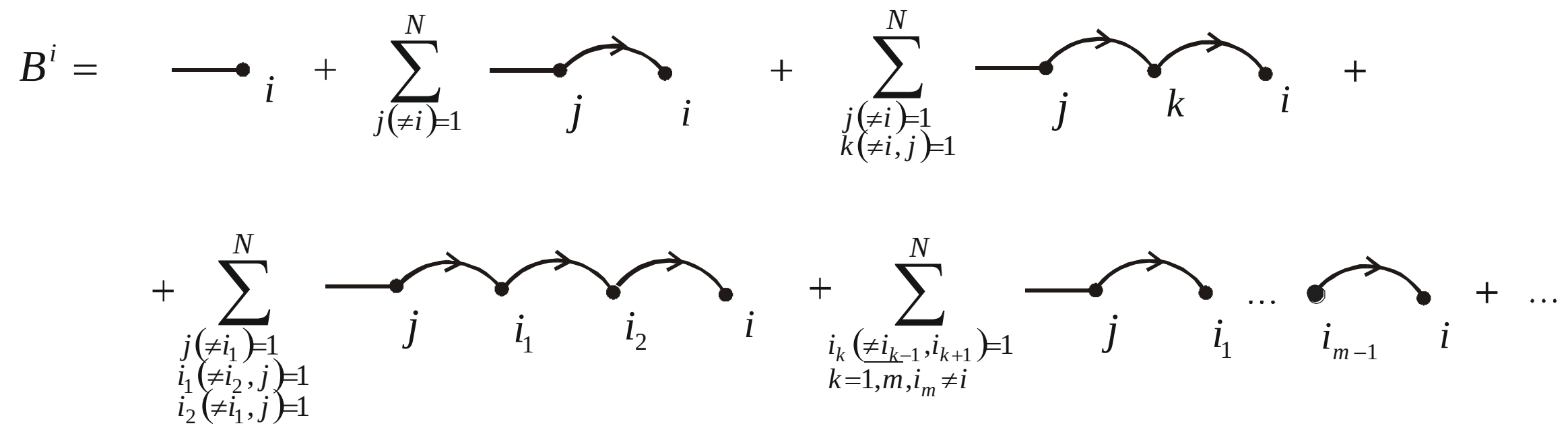

(50)

## 7. ТЕОРЕМА СЛОЖЕНИЯ ДЛЯ РЕГУЛЯРНЫХ ШАРОВЫХ ФУНКЦИЙ

Общая теорема сложения для регулярных шаровых функций выводилась различными способами многими авторами [35]. Самое простое доказательство этой теоремы было дано в статье [39].

Здесь мы покажем, что общая теорема сложения для регулярных шаровых функций, заданных в сферических координатах, является следствием известной теоремы Гобсона [40]. Имеет место следующий важный результат [40].

**Теорема 4.** (Гобсона). *Пусть* $p_n(\mathbf{x})$ *- однородный полином степени* $n$ *и рассмотрим сложную функцию* $F(\varphi(\mathbf{x})) \in C^n(\Omega)$, *где* $\varphi(\mathbf{x}) = |\mathbf{x}|^2$, *тогда выполняется соотношение*

$$p_n(\partial_{\mathbf{x}}) F(\varphi(\mathbf{x})) = \sum_{k=0}^{n} \frac{2^{n-2k}}{k!} \frac{d^{n-k}}{d\varphi^{n-k}} F(\varphi) \left(\nabla_{\mathbf{x}}^2\right)^k p_n(\mathbf{x}).$$

(51)

Из теоремы Гобсона непосредственно вытекает важное

**Следствие.** *Если* $F\left(|\mathbf{x}|^2\right) = 1/r$ $(r = |\mathbf{x}|)$, *то формула (51) может быть записана следующим образом:*

$$p_n(\partial_{\mathbf{x}}) \frac{1}{r} = (-1)^n \frac{(2n-1)!!}{r^{2n+1}} \left[1 - \frac{r^2}{2(2n-1)} \nabla^2 + \frac{r^4}{2 \cdot 4 \cdot (2n-1) \cdot (2n-3)} \nabla^4 - \ldots\right] p_n(\mathbf{x}).$$

(52)

Применяя формулу (52) для частного случая однородного многочлена $(x+iy)^m z^{l-m}$ степени $l$ и вычисляя производные по $z$ после перехода к сферическим координатам можно получить представление

$$\frac{1}{r^{l+1}}Y_{lm} = (-1)^l \frac{N_{lm}}{(l-m)!}\left(\partial_x + i\partial_y\right)^m \partial_z^{l-m}\left(\frac{1}{r}\right),$$

(53)

где $Y_{lm} = N_{lm}P_l^m(\mu)e^{im\phi}$, $\mu = \cos\theta$ и $N_{lm} = \left[(2l+1)(l-m)!/4\pi(l+m)!\right]^{1/2}$.

Формула (53) позволяет легко получить искомую теорему сложения.

**Теорема 5** (общая теорема сложения). *Для регулярных шаровых функций верно соотношение*

$$\frac{1}{r^{l+1}}Y_{lm}(\mathbf{r}) = \frac{4\pi N_{lm}}{(l-m)!}\sum_{l'=0}^{\infty}\sum_{m'=-l'}^{l'}\frac{(-1)^{l+m'}}{2l'+1}\frac{N_{l'm'}}{(l'-m')!}\frac{(l+l'-m+m')!}{N_{l+l',m-m'}}\times$$

$$\times\frac{r_1^{l'}}{L^{l+l'+1}}Y_{l+l',m-m'}(\mathbf{L})Y_{l'm'}(\mathbf{r}_1).$$

(54)

**Доказательство**. Предположим, что имеет место связь

$$r^2 = |\mathbf{r}_1 - \mathbf{L}|^2 = x^2 + y^2 + (z-L)^2 = r_1^2 + L^2 - 2r_1L\mu_1.$$

(55)

Для дальнейшего удобно использовать функции $Y_{l,-m}$, которые определяются как

$$Y_{l,-m} = (-1)^l \frac{N_{l,-m}}{(l+m)!}r^{l+1}\left(\partial_x + i\partial_y\right)^{-m}\partial_z^{l+m}\left(\frac{1}{r}\right),$$

(56)

причем хорошо известно, что

$$Y_{l,-m} = (-1)^m Y_{lm}^*.$$

(57)

Из уравнения Лапласа непосредственно следует, что

$$\partial_z^{l+m}\left(\frac{1}{r}\right) = (-1)^m\left(\partial_x + i\partial_y\right)^m\left(\partial_x - i\partial_y\right)^m\partial_z^{l-m}\left(\frac{1}{r}\right).$$

(58)

С помощью равенства (55) формулу (53) можно переписать в виде

$$\frac{1}{r^{l+1}}Y_{lm}(\mathbf{r}) = \frac{N_{lm}}{(l-m)!}\left(\partial_{L_x} + i\partial_{L_y}\right)^m\partial_{L_z}^{l-m}\left(\frac{1}{|\mathbf{r}_1 - \mathbf{L}|}\right).$$

(59)

Чтобы отсюда вывести общую теорему сложения мы здесь воспользуемся известным разложением фундаментального решения уравнения Лапласа [36]

$$\frac{1}{4\pi|\mathbf{r}_1 - \mathbf{L}|} = \sum_{l'=0}^{\infty}\sum_{m'=-l'}^{l'}\frac{1}{2l'+1}\frac{r_1^{l'}}{L^{l'+1}}Y_{l'm'}^*(\mathbf{L})Y_{l'm'}(\mathbf{r}_1) \text{ если } r_1 < L.$$

При помощи соотношений (56) и (57) это разложение может быть представлено как

$$\frac{1}{|\mathbf{r}_1 - \mathbf{L}|} = \sum_{l'=0}^{\infty}\sum_{m'=-l'}^{l'}(-1)^{l'+m'}\frac{4\pi}{2l'+1}r_1^{l'}Y_{l'm'}(\mathbf{r}_1)\frac{N_{l'm'}}{(l'-m')!}\left(\partial_{L_x} + i\partial_{L_y}\right)^{-m'}\partial_{L_z}^{l'+m'}\left(\frac{1}{L}\right).$$

(60)

Подстановка последнего разложения в равенство (59) после несложных преобразований дает искомую общую теорему сложения (54).

**Следствие.** В частном случае осевой симметрии (т.е. когда $m = 0$) из равенства (54) непосредственно следует осесимметричная теорема сложения

$$\frac{1}{r^{l+1}}P_l(\mu)=\sum_{n=0}^{\infty}(-1)^l\frac{(l+n)!}{l!n!}\frac{r_1^n}{L^{l+n+1}}P_n(\mu_1),$$

(61)

которая была выведена в классической монографии Гобсона [40].

## 8. ТЕОРЕМА СЛОЖЕНИЯ ДЛЯ ДЕКАРТОВЫХ НЕПРИВОДИМЫХ ТЕНЗОРОВ

Доказательство теоремы сложения для регулярных шаровых функций, заданных в сферической системе координат приведено выше. Здесь мы докажем теорему сложения для регулярных шаровых функций заданных при помощи декартовых неприводимых тензоров.

В общем случае неприводимый тензор $T_l$ ранга $l$ определяется как набор $2l+1$ функций $T_{lm}$ (где $m=\overline{-l,l}$), которые удовлетворяют следующим коммутационным правилам со сферическими компонентами оператора углового момента [41]

$$\left[\hat{J}_{\pm},T_{lm}\right]=\mp\frac{1}{\sqrt{2}}\sqrt{l(l+1)-m(m\pm1)}T_{lm\pm1},\quad \left[\hat{J}_z,T_{lm}\right]=mT_{lm}.$$

Можно показать, что соответствующий неприводимый декартовый тензор $k$-го ранга (напомним, что в декартовых координатах нет необходимости различать ковариантные и контравариантные координаты тензоров) определяется как [42]

$$\overline{x_{\gamma_1}...x_{\gamma_n}}=\frac{(-1)^n}{(2n-1)!!}r^{2n+1}\frac{\partial^n}{\partial x_{\gamma_1}...\partial x_{\gamma_n}}\left(\frac{1}{r}\right),$$

(62)

где $(2n-1)!!=1\cdot3\cdot...\cdot(2n-1)$, а $x_{\gamma_\nu}$ - декартовы координаты вектора $\mathbf{r}$, $\gamma_\nu=\overline{1,3}$. Неприводимые тензоры преобразуются по неприводимому представлению группы $SO_3$. По теореме Сильвестра [43] регулярные шаровые функции выражаются через декартовы неприводимые тензоры и наоборот. Другими словами неприводимые декартовы тензоры как и шаровые функции образуют базис в пространстве представления группы $SO_3$, поэтому существует невырожденное линейное преобразование в $\mathbf{R}^{2l+1}$ преобразующее эти базисы друг в друга.

В области $\Omega^*$ вне $N$ сфер рассмотрим локальные координаты совмещенные с центрами $i$-ой и $j$-ой сфер: $\mathbf{r}_i^0$ и $\mathbf{r}_j^0$. Мы будем использовать очевидную связь

$$\mathbf{L}_{ij}=\mathbf{r}_j^0-\mathbf{r}_i^0,\ \mathbf{r}_j=\mathbf{r}_i-\mathbf{L}_{ij},$$

(63)

где $\mathbf{r}_{i(j)}=\mathbf{r}-\mathbf{r}_{i(j)}^0$. Предположим, что декартовы координаты $j$-ой и $i$-ой сфер связаны лишь трансляциями, т.е. направления их координатных осей одинаковы. Тогда для декартовых координат $j$-ой и $i$-ой сфер получаем

$$x_{\gamma_\nu}^j=x_{\gamma_\nu}^i+a_{\gamma_\nu}^{ij},$$

(64)

где $a_{\gamma_\nu}^{ij}$ - некоторые постоянные. Имеет место

**Теорема 6** (теорема сложения). *Для неприводимых декартовых тензоров выполняется равенство*

$$r_j^{-(2n+1)}\overline{x_{\gamma_1}^j...x_{\gamma_n}^j}=\sum_{k=0}^{\infty}\omega_{kn}\left(\frac{R_j}{L_{ij}}\right)^{n+1}\left(\frac{R_i}{L_{ij}}\right)^k\Omega_{\gamma_1...\gamma_n\mu_1...\mu_k}\left(\mathbf{L}_{ij}\right)\overline{x_{\mu_1}^i...x_{\mu_k}^i},$$

(65)

*где* $\Omega_{\gamma_1...\gamma_n\mu_1...\mu_k}\left(\mathbf{L}_{ij}\right)=L_{ij}^{-(k+n)}\overline{L_{\gamma_1}^{ij}...L_{\gamma_n}^{ij}L_{\mu_1}^{ij}...L_{\mu_k}^{ij}}$,

$$\omega_{kn}=\frac{(-1)^n}{k!}\frac{\left[2(k+n)-1\right]!!}{(2n-1)!!}.$$

(66)

**Доказательство**. Линейная зависимость (64) приводит к очевидному соотношению

$$\frac{\partial}{\partial x_{\gamma_\nu}^j}=\frac{\partial}{\partial x_{\gamma_\nu}^i}.$$

(67)

Воспользовавшись этим соотношением в определении (62), а также связью (63) получаем

$$r_j^{-(2n+1)}\overline{x_{\gamma_1}^j...x_{\gamma_n}^j}=\frac{(-1)^n}{(2n-1)!!}\frac{\partial^n}{\partial x_{\gamma_1}^i...\partial x_{\gamma_n}^i}\left(\frac{1}{\left|\mathbf{r}_i-\mathbf{L}_{ij}\right|}\right)$$

или, принимая во внимание, что $\frac{\partial}{\partial x_{\gamma_\nu}^i}=-\frac{\partial}{\partial L_{\gamma_\nu}^{ij}}$, имеем

$$r_j^{-(2n+1)}\overline{x_{\gamma_1}^j...x_{\gamma_n}^j}=\frac{1}{(2n-1)!!}\frac{\partial^n}{\partial L_{\gamma_1}^{ij}...\partial L_{\gamma_n}^{ij}}\left(\frac{1}{\left|\mathbf{r}_i-\mathbf{L}_{ij}\right|}\right).$$

(68)

Подстановка известного разложения фундаментального решения [42]

$$\frac{1}{r_j}=\frac{1}{\left|\mathbf{r}_i-\mathbf{L}_{ij}\right|}=\sum_{k=0}^{\infty}\frac{(-1)^k}{k!}\frac{\partial^n}{\partial L_{\gamma_1}^{ij}...\partial L_{\gamma_k}^{ij}}\left(\frac{1}{L_{ij}}\right)\overline{x_{\gamma_1}^i...x_{\gamma_k}^i}$$

(69)

в формулу (68) с учетом равномерной сходимости ряда (69) дает искомую теорему сложения (65).

Применим доказанную теорему для решения задачи Дирихле (1)-(3). Представим решение $u(\mathbf{r})$ в окрестности $i$-го шара в виде суперпозиции

$$u=\sum_{i=1}^{N}u_i,$$

(70)

где $u_i$ - произвольная регулярная гармоническая функция на $D_i$, т.е.

$$u_i=\sum_{n=0}^{\infty}A_{\gamma_1...\gamma_n}^i\xi_i^{-(2n+1)}\overline{\xi_{\gamma_1}^i...\xi_{\gamma_n}^i},$$

(71)

где $\xi_i = r_i / R_i$, $\xi^i_{\gamma_\nu} = x^i_{\gamma_\nu} / R_i$, а $A^i_{\gamma_1 ... \gamma_n}$ - некоторые тензорные коэффициенты, определяемые при помощи краевых условий (2). Разлагая функцию $f_i$ в ряд по ограничению неприводимых тензоров на $\partial\Omega_i$

$$f_i = \sum_{n=0}^{\infty} \alpha^i_{\gamma_1 ... \gamma_n} \left( \overline{\xi^i_{\gamma_1} ... \xi^i_{\gamma_n}} \right) \Big|_{\partial\Omega_i}$$

и используя теорему сложения (65) в граничных условиях (2), находим БСЛАУ для определения неизвестных коэффициентов $A^i_{\gamma_1 ... \gamma_n}$:

$$A^i_{\gamma_1 ... \gamma_n} = \alpha^i_{\gamma_1 ... \gamma_n} - \sum_{j(\neq i)=1}^{N} \sum_{k=0}^{\infty} A^i_{\mu_1 ... \mu_k} \omega_{nk} \left( \frac{R_j}{L_{ij}} \right)^{k+1} \left( \frac{R_i}{L_{ij}} \right)^{n} \Omega_{\gamma_1 ... \gamma_n \mu_1 ... \mu_k} \left( \mathbf{L}_{ij} \right), \quad i = \overline{1, N}, \quad n = \overline{0, \infty}.$$

(72)

Решение БСЛАУ (72) и формулы (70), (71) дают решение поставленной задачи Дирихле (1)-(3). Важно отметить, что совершенно аналогично может быть также решена и соответствующая смешанная краевая задача Дирихле-Неймана.

Решение задачи Дирихле, немного отличающееся от рассмотренного выше, было предложено в работах [44,45], причем выражение (66) записано в них как $(-1)^k \omega_{kn}$ (эта опечатка была исправлена в последующей статье [46]). Позднее метод решения задачи изложенный в работе [45], включая указанную, а также ряд других опечаток, был полностью воспроизведен Красовитовым в диссертации [47] и статьях [48-50], но представлен там как оригинальный. Отметим, что помимо этого работы [47-50] содержат грубейшие математические ошибки даже в основных формулах (см. подробное обсуждение данного вопроса в статье [51]).

## 9. ДРУГИЕ МЕТОДЫ РЕШЕНИЯ ЗАДАЧИ ДИРИХЛЕ

Метод индуцированных сил (МИС) был впервые предложен для решения задачи о гидродинамическом взаимодействии $N$ сферических частиц в стоксовском потоке [52,53], а затем применен к решению задачи о диффузии в среде с $N$ неподвижными сферическими стоками [54]. Решение указанных задач производилось в пространстве фурье-образов, однако, в переходе к фурье-образам нет необходимости, так как задача может быть решена в пространстве оригиналов [55]. Последнее представляется важным, поскольку в этом случае тождественность МИС методу разложения решения в ряды по неприводимым тензорам становится более очевидной. В связи с этим мы здесь рассмотрим вариант МИС для пространства оригиналов.

В МИС краевая задача (1)-(3) в области $\Omega^*$ ($\bar{\Omega}_i \equiv \bar{B}_{R_i}$) заменяется уравнением

$$\nabla^2 u = -4\pi \sum_{i=1}^{N} \rho_i(\mathbf{r})$$

(73)

с теми же условиями (2) и (3). Функции $\rho_i$ в правой части уравнения (73) обычно определяются выражением

$$\rho_i(\mathbf{r}) = R_i^{-2} g_i(\xi_i) \delta\left( \left| \mathbf{r} - \mathbf{r}_i^0 \right| - R_i \right),$$

(74)

где $g_i \in C(\partial\Omega_i)$. Для простоты мы здесь положим, что $R_i = R$ $(i=\overline{1,N})$. Очевидно, что общее решение уравнения (73) дается квадратурой

$$u(\mathbf{r}) = \sum_{i=1}^{N} \int \frac{\rho_i(\mathbf{r}')d\mathbf{r}'}{|\mathbf{r}-\mathbf{r}'|}.$$

(75)

Воспользовавшись разложением (69) представим ядро интегрального оператора в (75) как

$$\frac{1}{|\mathbf{r}-\mathbf{r}'|} = \frac{1}{|\mathbf{r}_i-\mathbf{r}'_i|} = \sum_{l=0}^{\infty} \frac{(2l-1)!!}{l!} r_i^{-(l+1)} \overline{x^i_{\gamma_1}...x^i_{\gamma_l}}\,\overline{\xi^i_{\gamma_1}...\xi^i_{\gamma_l}},$$

(76)

где $\mathbf{r}'_i = \mathbf{r}' - \mathbf{r}_i^0$. Введем так называемый коннектор [55]

$$A^{(0,l)}(\mathbf{r}_i) = (2l-1)!!\left(\frac{R}{r_i}\right)^{l+1} \overline{\xi^i_{\gamma_1}...\xi^i_{\gamma_l}}$$

(77)

и поверхностные моменты плотности источников

$$\rho_i^{(l)}(\mathbf{r}_i) = \frac{1}{l!R^l} \int \overline{x'^i_{\gamma_1}...x'^i_{\gamma_l}}\,\rho_i(\mathbf{r}')d\mathbf{r}'.$$

(78)

Подставляя разложение (76) в формулу (75) и учитывая равномерную сходимость ряда (76) при $r'_i < r_i$, а также, используя определения (77) и (78), находим

$$u(\mathbf{r}) = \frac{1}{R} \sum_{i=1}^{N} \sum_{l=0}^{\infty} A^{(0,l)}(\mathbf{r}_i)\,\rho_i^{(l)}(\mathbf{r}_i).$$

(79)

Неизвестные поверхностные моменты плотности источников должны быть найдены с помощью граничных условий (2). Легко видеть, что полученное выражение с точностью до несущественных постоянных совпадает с выражением (70), (71). Более того, система уравнений для определения моментов плотности источников полностью совпадает с БСЛАУ (72). Отсюда очевидно, что в неявном виде МИС использует теорему сложения для декартовых неприводимых тензоров.

Для решения многочисленных задач, описывающих диффузионно-контролируемые процессы в системах с большим числом сферических идеальных стоков ($f_i \equiv 1$), Кукером и Фридом был предложен следующий подход [56]. Рассмотрим уравнение Пуассона вида

$$\nabla^2 u = \sum_{i=1}^{N} \int_{\partial\Omega_i} d\omega_i \delta(\mathbf{r}-\mathbf{r}_i)\sigma_i(\omega_i),$$

(80)

где $\sigma_i(\omega_i) \in C(\partial\Omega_i)$ - поверхностная плотность силы стока, $\mathbf{r}_i = \mathbf{r}_i^0 + \mathbf{r}_i(\omega_i) \in \partial\Omega_i$, а $\omega_i$ обозначает ориентацию вектора $\mathbf{r}_i$. Формальное решение уравнения (80) дается формулой

$$u(\mathbf{r}) = -\int_{\partial\Omega_i} d\omega_i G_0(\mathbf{r}-\mathbf{r}_i)\sigma_i(\omega_i) - \sum_{j(\neq i)=1}^{N} \int_{\partial\Omega_j} d\omega_j G_0(\mathbf{r}-\mathbf{r}_j)\sigma_j(\omega_j).$$

(81)

где $G_0(\mathbf{r}) = 1/4\pi|\mathbf{r}|$ - фундаментальное решение уравнения Лапласа.

Введя в рассмотрение ограничение функции $G_0$ на границу: $G_0(\omega_i, \omega_i') = G_0(\mathbf{r}_i - \mathbf{r}_i')\big|_{\partial\Omega_i}$ предполагается существование граничного обратного оператора $\mathbf{G}_i^{-1}$ такого, что

$$\int_{\partial\Omega_i} d\omega_i'' G_i^{-1}(\omega_i, \omega_i'') G_0(\omega_i'', \omega_i') = \delta(\omega_i - \omega_i').$$

(82)

Подстановка решения (81) в граничные условия (2) и применение оператора $\mathbf{G}_i^{-1}$ приводит к интегральным уравнениям Фредгольма II рода относительно функции $\sigma_i(\omega_i)$, причем в работе [56] полученные уравнения решаются методом итераций.

Из сравнения с МИС видно, что метод Кукера-Фрида разнится лишь правой частью в исходном уравнении (80), но в отличие от МИС он не использует теорему сложения и поэтому сводит рассматриваемую задачу к решению довольно громоздких интегральных уравнений.

Для приближенного решения внешних краевых задач диффузии в областях с многосвязной границей Бурлацкий и Овчинников предложили следующий метод [57]. Приближенная функция Грина задачи Дирихле представляется в виде

$$G\left(\mathbf{r}, \mathbf{r}'; \mathbf{r}_1^{(0)}, \ldots, \mathbf{r}_N^{(0)}\right) = \frac{\det \mathbf{F}}{\det \mathbf{H}},$$

(83)

где

$$\mathbf{F} = \begin{pmatrix} G_0(\mathbf{r} - \mathbf{r}') & G_0\left(\mathbf{r} - \mathbf{r}_1^{(0)}\right) & \cdots & G_0\left(\mathbf{r} - \mathbf{r}_N^{(0)}\right) \\ G_0\left(\mathbf{r}_1^{(0)} - \mathbf{r}'\right) & & & \\ \vdots & & \mathbf{H} & \\ G_0\left(\mathbf{r}_N^{(0)} - \mathbf{r}'\right) & & & \end{pmatrix}$$

и

$$\mathbf{H} = \begin{pmatrix} G_0(R) & G_0(\mathbf{L}_{21}) & \cdots & G_0(\mathbf{L}_{N1}) \\ G_0(\mathbf{L}_{12}) & G_0(R) & \cdots & G_0(\mathbf{L}_{N2}) \\ \vdots & \vdots & \ddots & \vdots \\ G_0(\mathbf{L}_{1N}) & G_0(\mathbf{L}_{2N}) & \cdots & G_0(R) \end{pmatrix}.$$

Легко видеть, что функция (83) является симметричной по переменным $\mathbf{r}$ и $\mathbf{r}'$, удовлетворяя при этом уравнению

$$-\nabla^2 G = \delta(\mathbf{r} - \mathbf{r}').$$

Таким образом, приближенное решение краевой задачи (1)-(3) дается квадратурой

$$u(\mathbf{r}) = \sum_{i=1}^{N} \int_{\partial\Omega_i} f_i(\omega_i) \frac{\partial}{\partial \mathbf{r}_i'} G\left(\mathbf{r}, \mathbf{r}'; \mathbf{r}_1^{(0)}, \ldots, \mathbf{r}_N^{(0)}\right) dS_i.$$

Важно отметить, что метод Бурлацкого-Овчинникова позволяет также приближенно решить соответствующую нестационарную задачу Дирихле [57]. Можно показать, что метод Бурлацкого-Овчинникова дает результаты эквивалентные монопольному приближению, т.е. приближению соответствующему системе уравнений (72) и общему решению (70), (71) при $n = 0$.

## 10. ВЫВОДЫ

В статье на примере внешней задачи Дирихле для уравнения Лапласа дан подробный анализ существующих аналитических методов решения краевых задач в областях с многосвязной границей. Для трехмерных областей вне непересекающихся шаров доказано достаточное и необходимое условия сходимости метода отражений (19) и (20). Показано, что метод изображений представляет собой частный случай метода отражений, что позволяет обойтись без использования теорем сложения. Кроме того, так как метод изображений использует инвариантность относительно преобразования Кельвина, он применим лишь при решении задач для уравнения Лапласа. Найдена связь метода отражений и прямого метода использующего теоремы сложения, причем процедура метода простых итераций для соответствующей бесконечной системы линейных алгебраических уравнений совпадает с итерациями метода отражений. Оказалось, что достаточное условие сходимости метода отражений (простых итераций) (19) слабее хорошо известного достаточного условия (48). Показано, что теорема сложения для регулярных шаровых функций является следствием известной теоремы Гобсона. Предложен метод решения внешней задачи Дирихле, основанный на теореме сложения для неприводимых декартовых тензоров. Показано, что в пространстве оригиналов метод индуцированных сил приводит к бесконечной системе линейных алгебраических уравнений совпадающей с той, которая получается при помощи теоремы сложения для неприводимых декартовых тензоров. Отмечено, что метод Кукера в отличие от метода индуцированных сил не использует теорему сложения, что сводит рассматриваемую задачу к решению довольно громоздких интегральных уравнений Фредгольма, а метод Бурлацкого-Овчинникова эквивалентен монопольному приближению.

*В заключении автор выражает свою глубокую благодарность проф.А.А.Юшканову за обсуждение результатов работы и полезные замечания.*

## ЛИТЕРАТУРА

1. Бажал И.Г., Куриленко О.Д. Перекристаллизация в дисперсных системах. Киев, Наукова думка, 1975, 216 с.
2. Brailsford A.D., Bullough R. The theory of sink strengths. Philos. Trans. R. Soc. London, Ser. A, 1981, v.302, p.87-137.
3. Кристенсен Р. Введение в механику композитов. М., Мир, 1982, 336 с.
4. Calef D. F., Deutch J. M. Diffusion- controlled reactions. Ann. Rev. Phys. Chem., 1983, v.34, p.493-524.
5. Крюкова Ю.Ю., Моденов В.П. Краевые задача для уравнения Гельмгольца в многосвязной волноводной области с кусочнопостоянной границей. Вестн. Моск. ун-та, сер. Физ. Астрон., 2002, №3, с.36-40.
6. Згаевский В.Э., Яновский Ю.Г. Вычисление эффективной вязкости концентрированных суспензий жестких частиц на основе кристаллической модели. Механика композиционных материалов и конструкций, 1996, т.2, №1, с.137-167.

7. Maury B. A. fat boundary method for the Poisson problem in a domain with holes. SIAM J. Sci. Comp., 2001, v.16, p.319-339.
8. Sabelfeld K.K., Simonov N.A. Random walks on boundary for solving PDEs. VSP, Utrecht, 1994, 133 p.
9. Tsao H.-K., Lu S.-Y., Tseeng C.-Y. Rate of diffusion-limited reactions in a cluster of spherical sinks. J. Chem. Phys., 2001, v.115, p.3827-3833.
10. Кручек М.П. Основы векторного и тензорного исчисления. Петрозаводск, 1983, 88 с.
11. Векуа И.Н. Новые методы решения эллиптических уравнений. Л., ОГИЗ, 1948, 296 с.
12. Axler S., Bourdon P., Ramey W. Theory of harmonic functions. New York, Springer-Verlag, 1992, 232 p.
13. Гюнтер Н.М. Теория потенциала и ее применение к основным задачам математической физики. М., Гостехиздат, 1953, 415 с.
14. Хенл Х., Мауэ А., Вестфаль К. Теория дифракции. М., 1964, 428 с.
15. Соболев С.Л. Алгоритм Шварца в теории упругости. Докл. АН СССР, 1936, т.4, №6, с.235-238.
16. Михлин С.Г. Линейные уравнения в частных производных. М., Высшая школа, 1980, 431 с.
17. Gonzalez R., Kress R. On the treatment of a Dirichlet-Neumann mixed boundary value problem for harmonic functions by an integral equation method. SIAM J. Math. Anal., 1977, v.8, p.504-517.
18. Крутицкий П.А. Смешанная задача для уравнения Лапласа в трехмерной многосвязной области страницы. Дифф. уравнения, 1999, т.35, с.1179-1186.
19. Иванов Е.А. Дифракция электромагнитных волн на двух телах. Минск, 1968, 589 с.
20. Агошков В.И., Дубовский П.Б., Шутяев В.П. Методы решения задач математической физики. М., Физматлит, 2002, 320 с.
21. Smoluchowski M.V. Uber die wechselwirkung von kugeln, die sich in einer žahen flussigkeit bewegen. Bull. Intern. Acad. Polonaise Sci. Lett., 1911, v.1A, p.28-39.
22. Tversky V. Multiple scattering of radiation by an arbitraryplanar configuration of paralle; cylinders and by two parallel cylinders. J. Appl. Phys., 1952, v.22, p.407-414.
23. Хаппель Дж., Бреннер Г. Гидродинамика при малых числах Рейнольдса. М., Мир, 1976, 630 с.
24. Ohshima H. Electrostatic interaction between two spherical colloidal particles Adv. Colloid Interfase Sci., 1994, v.53, p.77-102.
25. Traytak S.D., Tachiya M. Diffusion-controlled reactions in an electric field: Effects of an external boundary and competition between sinks. J. Chem. Phys., 1997, v.107, p.9907-9920.
26. Stimson M., Jeffry G.B. The motion of two spheres through a viscous fluid. Proc. Roy. Soc., 1926, v.A111, p.110-116.
27. Годунов С.К. Уравнения математической физики. М., Наука, 1971, 416 с.
28. Strieder W., Aris R. Variational methods applied to problems of diffusion and reaction. New York, Springer, 1973, 109 p.
29. Blawzdziewicz J., Szamel G., Van Beijeren H. Diffusion-controlled reactions: Upper bounds on the effective rate constant. J. Chem. Phys., 1991, v.94, p.7967-7971.

30. Голузин Г.М. Решение пространственной задачи Dirichlet для уравнения Laplace'a для областей, ограниченных конечным числом сфер. Матем. сб., 1934, т.41, №2, с.277-283.
31. Jeans J.H. The mathematical theory of electricity and magnetism. Cambridge, 1966, 652 p.
32. Франк Ф., Мизес Р. Дифференциальные и интегральные уравнения математической физики. Л.-М., ОНТИ, 1937, 998 с.
33. Доценко И.С. Решение уравнения Лапласа в пространстве, ограниченном двумя неконцентрическими сферами. Вестн. Киев. Ун-та, 1987, т.28, с.36-39.
34. Миллер У. Симметрия и разделение переменных. М., Мир, 1981, 343 с.
35. Weniger E.J. Addition theorems as three-dimensional Taylor expansions. Int. J. Quantum Chem., 2000, v.76, p.280-295.
36. Арфкен Г. Математические методы в физике. М., Атомиздат, 1970, 712 с.
37. Гайдуков М.Н., Мовсисян Л.М., Терзян А.В. Движение многих сферических частиц при наличии гидродинамического взаимодействия между ними в неизотермической вязкой среде. "Избранные вопросы теоретической и математической физики", М., 1987, деп. в ВИНИТИ № 6584-В87, с.60-91.
38. Канторович Л.В., Крылов В.И. Приближенные методы высшего анализа. Л., Физматгиз, 1962, 708 с.
39. Caola M.J. Solid harmonics and their addition theorems. J. Phys. A Math. Gen., 1978, v.11, p.123-125.
40. Гобсон Е.В. Теория сферических и эллипсоидальных функций. М., Изд-во иностр. лит., 1952, 476 с.
41. Варшалович Д.А., Москалев А.Н., Херсонский В.К. Квантовая теория углового момента. Л., Наука, 1975, 440 с.
42. Hess S., Kohler W. Formeln zur Tensor-Rechnung. Erlangen, Palm and Enke, 1980, 41 p.
43. Курант Р., Гильберт Д. Методы математической физики, т.I, М., Гостехиздат, 1933, 525 с.
44. Трайтак С.Д. Теория квазистационарной переконденсации капель. XV Всесоюзная конференция “Актуальные вопросы физики аэродисперсных систем” тезисы, 1989, Одесса, т.1, с.30.
45. Трайтак С.Д. К теории переконденсации $N$ капель. ТОХТ, 1990, т.24, с.473-482.
46. Traytak S.D. The diffusive interaction in diffusion-limited reactions: the steady-state case. Chem. Phys. Lett., 1992, v.197, p.247-254.
47. Красовитов Б.Г. Испарение и конденсационный рост крупных и умеренно крупных капель в газообразных средах при произвольных перепадах температуры. Диссертация канд. физ-мат наук, М., 1991, 134 с.
48. Elperin T., Krasovitov B. Combustion of cylindrical and spherical random clusters of char/carbon particles. Combust. Sci. and Tech., 1994, v.102, p.165-180.
49. Elperin T., Krasovitov B. Analysis of evaporation and conbustion of random clusters of droplets by a modified method of expansion into irreducible multipoles. Atomization and Sprays, 1994, v.4, p.79-97.
50. Elperin T., Krasovitov B. Evaporation and growth of multicomponent droplets in random dence clusters. Trans. ASME, 1997, v.119, p.288-297.
51. Traytak S. D. On the irreducible tensors method in the theory of diffusive interaction between particles. В кн.: Mathematical Modeling: Problems, methods, applications, Kluwer, 2001, c.267-278.

52. Mazur P., van Saarlos W. Many-sphere hydrodynamic interactions and mobilities in a suspension. Physica, 1982, v.115A, p.21-57.
53. Кульбицкий Ю.Н. О гидродинамическом взаимодействии частиц в приближении уравнений Стокса. М., Препринт СМНС АН СССР, №15, 1986, 64 с.
54. Beenakker C.W.J., Ross J. Monte Carlo study of a model of diffusion-controlled reactions. J. Chem. Phys., 1986, v.84, p.3857-3864.
55. Geigenmüller U., Mazur P. The effective dielectric constant of a dispersion of spheres. Physica A, 1986, v.136, p.316-369.
56. Cukier R.I., Freed K.F. Diffusion controlled processes among stationary reactive sinks: Effective medium approach. J. Chem. Phys., 1983, v.78, p.2573-2578.
57. Бурлацкий С.Ф., Овчинников А.А. Влияние флуктуаций плотности реагентов на кинетику процессов рекомбинации, размножения и гибели. ЖЭТФ, 1987, т.92, с.1618-1635.